# Rifrazione differenziale, tempo del solstizio invernale 2017 e obliquità vera dell'eclittica misurate alla Meridiana di Santa Maria degli Angeli in Roma

## Costantino Sigismondi (ICRA/Sapienza e Liceo G. Ferraris, Roma)

### Riassunto:

I dati del solstizio invernale 2017 presi a Santa Maria degli Angeli in Roma sulla linea meridiana di Francesco Bianchini del 1702 vengono trasformati in altezza del centro del Sole, corretti per l'effetto della rifrazione atmosferica ed utilizzati per calcolare l'istante del solstizio invernale e l'obliquità vera dell'eclittica.
Nel video https://youtu.be/Fi7P0ez7Ea4 si vede la misura di uno dei giorni, il 22 dicembre.
Come misurare l'istante del solstizio quando questo non capita al meridiano? Facendo un fit parabolico dei dati.
Già Giandomenico Cassini nel 1655 in san Petronio a Bologna fece vedere che in declinazione il Sole compie un moto oscillatorio tra -23.5° e +23.5°, nell'immagine (Cassini, 1695) ciò è rappresentato con un epiciclo, da cui partono le proiezioni di tutti i segni zodiacali sulla meridiana che sta sul pavimento della chiesa.

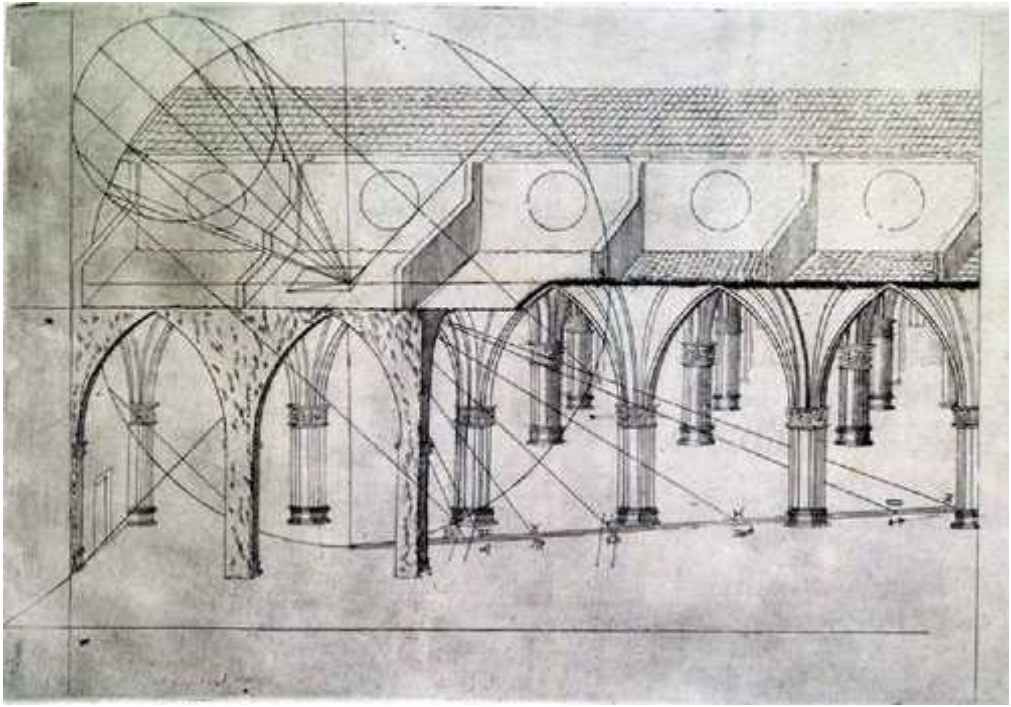

Il moto oscillatorio è il risultato della proiezione di un moto circolare
Il moto oscillatorio è armonico e ubbidisce alla legge di Hooke; attorno al solstizio l'accelerazione è costante fintanto che la declinazione (angolo, la nostra variabile spaziale) è stazionaria attorno al massimo (o minimo) valore.
Se l'accelerazione è costante, anche con una meridiana non esattamente tarata, basta misurare le differenze con un istante di riferimento (t=0 per noi sarà il mezzodi' del 21 dicembre) a t=1,2,3,4, e 8 e verificare che crescano quadraticamente; poi dalle equazioni della parabola in quei punti si ricavano le tre incognite della parabola: ampiezza, tempo del minimo e quota minima. L'esempio fatto nel testo risulta molto semplice, e basato su una matematica elementare, molto più di quella necessaria per capire i fit quadratici.
Il tempo del minimo da' il solstizio con meno di un'ora di errore (Bianchini, che costrui' la meridiana dava i risultati con meno errore, ma comparava il transito solare con quello delle stelle osservate dalla stessa meridiana, cosa che oggi non si puo' fare più).
La quota minima permette di ricavare la declinazione del Sole all'istante del Solstizio e quindi calcolare l'obliquità vera dell'eclittica al meglio di 4 secondi d'arco di precisione (dovuti alla turbolenza atmosferica e al contrasto variabile dell'immagine da un giorno all'altro).
L'obliquità vera è l'inclinazione istantanea dell'asse terrestre sul piano dell'orbita, addizionata della nutazione il cui valore è compreso tra + e -9 secondi d'arco e che fu scoperta con misure stellari nel 1737 dall'astronomo reale James Bradley.
Le misure accurate al millimetro fatte a santa Maria degli Angeli sulla meridiana del 1702 consentono precisioni finali di qualche secondo d'arco, semplicemente applicando la correzione cassiniana alle misure dirette decurtate della penombra (pari al diametro del foro stenopeico che è l'obbiettivo della meridiana).
Il solstizio di inverno fornisce le migliori circostanze per misurare i parametri dell'obliquità dell'eclittica come al video (2016)
https://www.youtube.com/watch?v=NiQjIt7fAF0

**Abstract and "pipeline"**

The declination of the Sun along the year varies according to a sinusoid. Around the solstices this curve is approximated by a parabola. In kinematics a parabola is obtained with a constant acceleration. This acceleration has been estimated in the days 21-29 December 2017, from the measurements taken at the meridian line of Santa Maria degli Angeli in Rome made by Francesco Bianchini in 1702 with purpose of measuring the variation of the obliquity of the ecliptic. Fitting the parabola equation to the data allowed to obtain also the instant of the solstice with an accuracy of one hour with respect of the ephemerides of IMCCE, used also to evaluate the accuracy of the measurements held at the meridian line within 3 arcseconds, the value of the seeing during the solar transits.
Being horizontal the meridian line and the pinhole, on the brass line are indicated the values of the tangent of the zenital angle multiplied by 100, so that each step corresponds to one hundredth of the height of the pinhole, and it is possible to read directly the tan(z) of the Sun on the meridian line.
In the paper is described the pipeline of the algorithm used to obtain the angular data of the center of the Sun starting from the ground measurements affected by the atmosphere:
1. reading of the observed tangents tan(z) of the zenital angles of the Southern and of the Northern limb of the Sun.
2. subtraction or addiction of the penumbra equivalent to half of the diameter of the pinhole, restored to the original dimension of 21 mm for this occasion, respectively for the Southern and for the Northern limb.
3. Calculation of the observed zenital angles z inverting their tangents
4. Application of the Cassini correction to the calculated zenital angles $z^*=z-60"*\tan(z)$
5. Calculation of the zenital angles $z^*$ of the center of the Sun by averaging the Southern and the Northern limb $z^*$.
6. Comparison of these declination with IMCCE ephemerides calculated for the pinhole's longitude and latitude (topocentric data) to evaluate the accuracy of the measurements.
7. The daily differences D in arcseconds with the zero reference on the 21st December transit (at 12:08) show the evolution of the parabola $D=a*(t-s)^2+d$ at t=0,1,2,3,4 and 8 days, and the solution of the six equations in 3 unknowns gives the exact time of the minimum "s", the value of the parabola's concavity or the angular acceleration "a", and the declination difference "d" between the 21st December transit and the instant of the solstice "s".
Since the equations are redundant with respect to the unknowns it is possible also to evaluate the errors associated to the solutions s, a and d. For the solstice timing s the errorbar is 2 hours and its value in good agreement with the ephemerides, and d allows to measure the instant value of the obliquity (composed with the nutation) within 4 arcseconds of accuracy. The relatively large time error on the solstice instant is due to the very slow motion in declination of the Sun in the hours around it, which becomes smaller than the observational angular resolution, limited by the atmospheric turbulence and the luminous contrast of the image in the church.
Francesco Bianchini in 1703 reduced the error on the solstices timings by using the measurement of the difference in right ascension of the Sun and of a star observed at the same meridian line even in daytime (as he did with Sirius in June-July 1703). The present one is an absolute measurement without stellar references.
 The meridian diameter is averagely measured 24 arcsecond less than the true value, and its standard deviation is 20 arcsec, because of different luminosity contrasts in the sky and in the Basilica, while the center of the image is much better defined (the contrast acts symmetrically without moving the center of the image), allowing an accuracy to the nearest arcsecond in the determination of the true obliquity.

**Introduzione**

Attorno al solstizio d'inverno l'immagine del Sole sulla meridiana di Francesco Bianchini nella Basilica di S. Maria degli Angeli raggiunge le dimensioni di 110 cm ed 1 mm corrisponde a circa 2". La turbolenza atmosferica fa tremare l'immagine per quantità maggiori del millimetro ed il contrasto luminoso all'interno della Basilica, che nel '700 veniva oscurata durante le osservazioni astronomiche, fa perdere parte della penombra attorno all'immagine.
Entrambi gli effetti sono casuali, il primo dipende dalle condizioni dell'alta atmosfera e da quelle sul piano dove si trova il foro stenopeico, il secondo dalla luminosità del cielo con o senza nubi, e dall'illuminazione della chiesa in caso di celebrazioni.
Esaminiamo qui i sei dati del solstizio invernale del 2017, dal 21 al 29 dicembre, presi al mezzogiorno locale, con l'apertura del foro stenopeico riportata a 21 mm di diametro, secondo il progetto di Bianchini che lo voleva 1/1000 dell'altezza di 20.34 m.
E' la prima volta, documentata, a partire dal 1750, che la meridiana è illuminata da un foro dalle dimensioni originali; negli ultimi 64 anni il foro è stato di dimensioni inferiori e in qualche tempo anche di forma non circolare (Catamo e Lucarini 2011).
Al fine di calibrare la posizione del nuovo foro ripristinato il 20 dicembre 2017 (video della presentazione della modifica https://youtu.be/IKmOBj0kxgY ) sono state prese misure accurate dell'immagine solare al solstizio e nei giorni successivi.

**Dati del 2017**

I dati solstiziali delle due estremità dell'immagine solare sono riferiti alle giunzioni tra le barre di ottone che compongono la linea meridiana vera e propria, in corrispondenza dei numeri 220 e 215. Queste giunzioni non sono molto nette, come invece sono i segni intermedi (come per i numeri 214, 213...). Su queste giunzioni sono stati fissati con dello scotch di carta dei ritagli di carta su cui ho segnato i limiti dei bordi solari meridionale e settentrionale. Al termine del passaggio meridiano è stata misurata direttamente anche la dimensione longitudinale dell'immagine solare.
Il video dell'osservazione del 22 dicembre è disponibile al sito https://youtu.be/Fi7P0ez7Ea4
Il 29 dicembre si è potuto stabilire che le posizioni del transito misurate anche ai bordi del marmo bianco circa 45 cm a destra e a sinistra della linea meridiana sono consistenti con quella sulla linea meridiana entro 1 mm, e quindi possono sostituire utilmente quelle meridiane in caso di nuvole.
Quando possibile anche una misura del tempo dei contatti tra l'immagine ed il centro della meridiana è stata presa, ai fini della taratura lungo la direzione Est-Ovest della posizione del foro stenopeico.

| Data | Posizione del Bordo Meridionale rispetto al 220 | | Posizione del Bordo Settentrionale rispetto al 215 | tempo passaggio |
|---|---|---|---|---|
| 18 Dicembre* | -38.5 mm | centro: 217-27mm | -134.5 mm | 12:07:01 |
| 21 Dicembre | +28 mm | | -46 mm | 12:08:33.5 |
| 22 Dicembre | +32.5 mm | | -54 mm | 12:09:05.5* |
| 23 Dicembre | +11.5 mm | | -75.5 mm | 12:09:33.5 |
| 24 Dicembre | -28 mm | | -126 mm | 12:10:07 |
| 25 Dicembre | -83 mm | | -164 mm | 12:10:35 |
| 29 Dicembre | -449.5 mm (217+158 mm**) | | (213-108mm**) -512.5 mm | 12:12:31.5 |

* il foro era in altra posizione e di altra dimensione
**prese in corrispondenza delle tacche originali, più definite e precise rispetto al 220 e al 215

### Prima operazione: "dempta penumbra"

Per un foro stenopeico vale, in ottica geometrica, l'analogo del principio di Huygens-Fresnel in ottica ondulatoria: ogni punto del foro stenopeico genera un'immagine del Sole sullo schermo. Per un foro stenopeico circolare l'immagine solare è più grande di quella che sarebbe generata da un obbiettivo puntiforme esattamente del diametro del foro stenopeico, se questo è posto in orizzontale e l'immagine anche'essa è orizzontale.
Fori stenopeici casuali di forma non circolare generano immagini solari che sono la convoluzione di tante immagini elementari con la forma del foro. Per immagine elementare si intende quella formata dal foro puntiforme: quella perpendicolare alla direzione dei raggi solari ha diametro $D=F*\tan(\theta)$. con $\theta$ diametro angolare del Sole in quel momento.
Quindi con un foro circolare la prima operazione da fare sui dati è sottrarre la penombra, metà per il bordo meridionale, che è il più lontano dal foro stenopeico, e metà per l'altro, con l'attenzione che qui il bordo teorico si avvicina al 215.

| Data | Posizione del Bordo Meridionale rispetto al 220 | | Posizione del Bordo Settentrionale rispetto al 215 | tempo passaggio |
|---|---|---|---|---|
| 18 Dicembre* | -46.5 mm | centro: 217-27mm | -126.5 mm | 12:07:01 |
| 21 Dicembre | +17.5 mm | | -35.5 mm | 12:08:33.5 |
| 22 Dicembre | +22.0 mm | | -43.5 mm | 12:09:05.5* |
| 23 Dicembre | +1.0 mm | | -65.0 mm | 12:09:33.5 |
| 24 Dicembre | -38.5 mm | | -115.5 mm | 12:10:07 |
| 25 Dicembre | -93.5 mm | | -153.5 mm | 12:10:35 |
| 29 Dicembre | -460 mm | | -502.0 mm | 12:12:31.5 |

### Trasformazione delle posizioni nella corrispondente tangente dell'angolo zenitale (moltiplicata 100)

| Data | Posizione del Bordo Meridionale rispetto al 220 | | Posizione del Bordo Settentrionale rispetto al 215 | tempo passaggio |
|---|---|---|---|---|
| 18 Dicembre* | 219,771 | centro: 216,867 | 214,378 | 12:07:01 |
| 21 Dicembre | 220,086 | | 214,826 | 12:08:33.5 |
| 22 Dicembre | 220,108 | | 214,786 | 12:09:05.5 |
| 23 Dicembre | 220,005 | | 214,681 | 12:09:33.5 |
| 24 Dicembre | 219,811 | | 214,432 | 12:10:07 |
| 25 Dicembre | 219,540 | | 214,246 | 12:10:35 |
| 29 Dicembre | 217,725 | | 212,521 | 12:12:31.5 |

*foro in altra posizione

### Secondo passo: Cassini o Laplace? scelta per il primo ordine ed inversione dell'equazione

La rifrazione atmosferica sposta in alto le stelle di una quantità proporzionale alla tangente del loro angolo zenitale z.
Giandomenico Cassini con l'Eliometro di san Petronio nel 1655 trovò che ciò valeva 60"*tan(z).
Sulla meridiana di Santa Maria degli Angeli i numeri da 37 a 220 scritti sulla sinistra per chi guarda verso il foro stenopeico sono proprio 100 volte la tangente del corrispondente angolo z, scritto sulla destra.
Quindi la correzione "cassiniana" vale circa 2.2*60"=132" per il bordo meridionale e poco meno di 2.15*60"=126" per il bordo settentrionale.
Di conseguenza per effetto della rifrazione atmosferica il disco del Sole risulta schiacciato di 6" lungo la verticale, rispetto al diametro orizzontale che vale 1950.89" nel periodo di perielio (che cade il 4 gennaio) e che non subisce rifrazione percepibile (in realtà la rifrazione orizzontale vale 0.56" ed è quella che si produrrebbe allo zenith, poiché il Sole ha una dimensione finita e la rifrazione che sarebbe nulla allo zenith z=0 agisce sui bordi del Sole posti a z=16' dallo zenith).
Per chiarezza ribadisco che il Sole al meridiano il giorno del Solstizio a Roma ha un diametro verticale di 1944.89" mentre quello orizzontale vale 1950.33".
L'effetto di rifrazione atmosferica aumenta all'aumentare dell'angolo zenitale, diventando massimo all'orizzonte (tramonto o alba)

con il massimo schiacciamento osservato del disco solare.
Pierre Simon de Laplace fece una trattazione analitica (il dato di Cassini era sperimentale, oggi lo chiameremmo un "fit" dei dati) e trovò che la formula più appropriata includeva anche un termine in tan^3(z), molto piccolo ma efficace per angoli zenitali maggiori. Il coefficiente teorico -0.067" che moltiplica il cubo della tangente nella formula di Laplace (Sigismondi, 2016) fa sì che per i nostri dati la correzione di Laplace valga circa 2.2^3*0.067=0.71" per il bordo meridionale e 0.66" per quello settentrionale. Questa correzione può essere trascurata dato che le frazioni di secondo d'arco non sono misurabili alla meridiana di Bianchini, ed il loro effetto differenziale non supera i 5/100 di secondo d'arco.
Più influente del termine in tan^3 è sicuramente il coefficiente della tangente nella formula di Cassini, che da valutazioni precedenti per Santa Maria degli Angeli (Sigismondi, 2016) vale 52.26" ed in letteratura per atmosfera standard vale 58.16" (Duffet-Smith, 1983).

Scegliamo dunque di usare la correzione cassiniana, come faceva Francesco Bianchini (1703) che costruì la meridiana per volontà del papa Clemente XI, usando sia 52" (vicino anche al valore invernale misurato a Pulkowo di 55") sia 60" di Cassini. Il termine di Cassini risulta il migliore.

**Terzo passo: l'arcotangente ed il calcolo del centro del Sole**

L'angolo zenitale z* che una stella puntiforme avrebbe senza la rifrazione dell'atmosfera viene spostato in alto di Δz=60"*tan(z*). Quindi l'angolo a cui osserviamo il nostro astro deve essere abbassato della stessa quantità.
z_oss=arctan(z) [tasti "inv" o "^-1" e funzione tan sulla calcolatrice scientifica] e z*=z_oss+60"tan(z_oss), che in gradi vale z*=z_oss+tan(z_oss)/60.
in questa ultima operazione ho considerato uguali tan(z*)=tan(z_oss), e ciò è vero entro due parti su mille, infatti il 18 dicembre abbiamo 2,19771 e tan(z*)=2,20144 con differenza di 373 parti su duecentoventimila, cioè 1.7 parti su mille.

| Data | z* del Bordo Meridionale | z* del centro del Sole (calcolato) | z* del Bordo Settentrionale | |
|---|---|---|---|---|
| 18 Dicembre* | 65,57019 | 65,29924 | 65,02829 | 12:07:01 |
| 21 Dicembre | 65,60116 | 65,33766 | 65,07416 | 12:08:33.5 |
| 21 Dicembre | 65,59627 | 65,33283 | 65,06938 | |
| 22 Dicembre | 65,60332 | 65,33669 | 65,07007 | 12:09:05.5* |
| 23 Dicembre | 65,59320 | 65,32626 | 65,05933 | 12:09:33.5 |
| 24 Dicembre | 65,57412 | 65,30397 | 65,03383 | 12:10:07 |
| 25 Dicembre | 65,54743 | 65,28109 | 65,01474 | 12:10:35 |
| 29 Dicembre | 65,367205 | 65,101829 | 64,837016 | 12:12:31.5 |

*foro obiettivo in altra posizione

Si può notare come i dati del 22 dicembre mostrino un bordo meridionale più basso di quelli del 21 dicembre (giorno in cui il solstizio è avvenuto alle ore 17:27), ciò è dovuto ad un diverso contrasto delle immagini, più marcato per il 22. La media tra distanza zenitale del bordo meridionale e di quello settentrionale corretti per effetto della rifrazione risulta comunque maggiore nel giorno del solstizio rispetto agli altri giorni.
La rifrazione differenziale come schiacciamento verticale dell'immagine solare (differenza tra la distanza zenitale dei bordi Nord e Sud) non è misurabile: i diametri risultano 1920" con deviazione standard di 20", sistematicamente inferiori di 24" al valore reale di 1944". Ciò è dovuto alle diverse condizioni di illuminazione della chiesa al momento delle misure. La misura del diametro orizzontale può essere fatta dal video https://youtu.be/guiL0LRh558 (di Daniele Impellizzeri) il transito del 22 dicembre è stato valutato in 2 min 22 s; la precisione di 1 secondo corrisponde a 1951"/142=13.7". In secondi d'arco il diametro orizzontale così misurato vale 142*15"/s*cos(23,4348)=1954,3" a cui però dovrebbe essere sottratta anche la penombra p=arctan(21mm/48757mm)=88.8". Il diametro verticale risulta molto più vicino al valore reale poiché è il risultato di due misure di posizione che durano un paio di secondi e nelle quali si possono mediare a occhio le oscillazioni della posizione del bordo del Sole, prendendo i valori massimi che includono la penombra. Il singolo contatto del bordo del Sole con la linea meridiana di ottone restituisce meno luce e quindi riduce il diametro percepito.

**Discussione: confronto con le effemeridi IMCCE ed incertezza nelle misure differenziali**

Se prendiamo come zero la posizione del centro del Sole al meridiano del 21 dicembre vediamo quanto si è alzato in declinazione (o equivalentemente in angolo zenitale) da un giorno all'altro.
Questo calcolo poteva essere fatto anche per ciascun bordo, ma l'effetto del contrasto varia da un passaggio all'altro ed il calcolo sul centro del Sole, come abbiamo visto adesso, dà risultati migliori.
Il confronto con le effemeridi IMCCE dell'Institut de Mécanique Céleste et Calcul des Ephémérides per il passaggio al meridiano di quei giorni fornisce accordi entro 3-4 secondi d'arco, fatta eccezione la misura del 24 dicembre, presa molto rapidamente durante la celebrazione della s. Messa, e che si scosta di 13 secondi d'arco. Le oscillazioni di 3-4 secondi d'arco attorno al valore delle effemeridi sono perfettamente consistenti con la grandezza della turbolenza atmosferica misurata in quei giorni sfruttando i passaggi su righe parallele dell'immagine solare della lente solstiziale di Salvador Cuevas nella cupola "Divinity in Light" (Cuevas e Sigismondi, 2016).

| Data | z* del centro del Sole | Δz* del centro del Sole in arcsec ["] | Δz* da effemeridi IMCCE del centro del Sole ["] |
|---|---|---|---|
| 18 Dicembre* | 65,29924 | +138.3 | +146.5 |
| 21 Dicembre | 65,33766 | +0.0 | + 0.0 |
| 22 Dicembre | 65,33669 | +3.5 | + 7.9 |
| 23 Dicembre | 65,32626 | +41.0 | +43.9 |
| 24 Dicembre | 65,30397 | +121.3 | +108.0 |
| 25 Dicembre | 65,28109 | +203.7 | +200.5 |
| 29 Dicembre | 65,101829 | +849.0 | +852.1 |

*foro obiettivo in altra posizione

**La calibrazione del nuovo foro stenopeico da effemeridi**

L'esame dei dati differenziali prescinde da eventuali errori di calibrazione del foro, quello che interessa è vedere se si riescono ad apprezzare le differenze di posizione del centro del Sole da un giorno all'altro e con quale affidabilità.
Abbiamo visto che per le osservazioni del 2017, durante 5 giorni di tempo sostanzialmente sereno dal 21 al 25 dicembre e poi il 29 dicembre, una precisione di 3-4 secondi d'arco è stata possibile. Giornate di atmosfera più stabile possono fornire condizioni in cui questi valori potrebbero addirittura dimezzare come si è già verificato in passato (Sigismondi, 2009).
Se invece confrontiamo gli angoli z* del centro del Sole con quelli calcolati dalle effemeridi, che si assumono implicitamente esenti da errori per lo meno entro 0.1", al fine di poter realizzare queste stime di accuratezza e le calibrazioni, troviamo la differenza sistematica di cui è necessario spostare il foro stenopeico nella direzione Nord-Sud per calibrarlo al meglio con le tacche sulla linea meridiana.
Infatti il foro non si trova più nella posizione originale in seguito a vari interventi, non documentati, che hanno modificato la piattaforma in cui era allocato, la sua posizione e finanche la sua forma.
Dal momento che la meridiana è stata scoperta disallineata rispetto al Nord-Sud di 4' 28.6" verso Est, questo comporta un ritardo sistematico dei tempi di passaggio dell'immagine solare rispetto alle effemeridi per quel foro stenopeico.
Questa situazione forse ha spinto qualcuno a realizzare un foro a forma di fagiolo, orientato grosso modo Est-Ovest, più piccolo del foro originale, probabilmente con l'intento di spostare a destra o sinistra una maschera circolare che fungeva da foro stenopeico, per correggere un poco il disallineamento di costruzione.
Ruggero Giuseppe Boscovich aveva misurato nel 1750 in 4'30" questo disallineamento con un ritardo di 17 secondi al solstizio invernale.
Attualmente vediamo un ritardo di 22 secondi, con una differenza di 5 secondi rispetto a Boscovich, che alla velocità dell'immagine di 3 mm/s corrispondono a 15-16 mm. Spostando il foro verso Est di 15-16 mm si riporta la meridiana nelle condizioni riscontrate da Boscovich, che sono molto verosimilmente quelle di costruzione (1702).

Per valutare di quanto occorre spostare nella direzione Nord Sud il foro stenopeico attuale, quello che ha fornito le misure in questo articolo dal 21 al 25 dicembre 2017, confrontiamo le posizioni teoriche del centro del Sole con quelle calcolate dalle nostre misure dopo la correzione cassiniana, e ne calcoliamo le tangenti per vederne la differenza.

| Data | z* calcolato del centro del Sole | z* da effemeridi IMCCE | Δtan(z) di cui spostare il foro + verso Nord |
|---|---|---|---|
| 21 Dicembre | 65,33766 | 65,3393 | + 0.016 |
| 22 Dicembre | 65,33669 | 65,3371 | + 0.004 |
| 23 Dicembre | 65,32626 | 65,3271 | + 0.009 |
| 24 Dicembre | 65,30397 | 65,3093 | + 0.014 |
| 25 Dicembre | 65,28109 | 65,2836 | + 0.025 |
| 29 Dicembre | 65,10183 | 65,1026 | + 0.008 |

la media degli spostamenti, da fare verso Nord, dell'attuale foro stenopeico, è di 12.7 parti millesime, cioè 2.6 mm con un'incertezza di 1.5 mm (7 parti millesime è la deviazione standard di questi spostamenti).

**L'algoritmo di Bianchini**

L'astronomo e numismatico Francesco Bianchini, che costruì la meridiana, misurava direttamente le parti millesime della centesima parte dell'altezza del foro stenopeico, alle quali si osservava la posizione dei lembi meridionale e settentrionale del Sole.
L'operazione "dempta penumbra" consisteva di levare 50 parti millesime al risultato osservato per il lembo meridionale ed aggiungerne 50 per quello settentrionale: una semplice sottrazione/addizione.
Poi la correzione cassiniana serviva ad abbassare le posizioni osservate riportandole al valore senza atmosfera. Per questa aggiungeva 1/60 della tangente letta sulla meridiana per ciascun lembo, poi Bianchini invertiva le tangenti per trovare gli archi corrispondenti.
Il centro del Sole era dato dalla media dei due lembi.
Nel caso di transiti tra il solstizio estivo e gli equinozi si poteva agevolmente misurare il punto in cui l'immagine toccava la meridiana e in cui l'abbandonava per valutare direttamente la posizione del centro del Sole senza ulteriori calcoli di media, ma solo con la correzione cassiniana, e non serviva sottrarre la penombra.
L'operazione di sottrazione della penombra poteva essere saltata nel caso di calcolo della posizione del centro del Sole, ma non poteva essere saltata se si voleva usare il lembo del Sole come indicatore della declinazione del centro, come per gli equinozi.
Per questo Bianchini aveva messo a punto il cronometro degli equinozi (Catamo e Lucarini 2011) per valutare dalla posizione osservata del lembo dell'immagine solare il tempo trascorso o mancante all'equinozio.

## Il moto armonico del Sole, l'istante del Solstizio e l'Obliquità dell'eclittica

Era ben noto ai greci (Tolomeo di Alessandria), disegnato da Cassini nello schema costruttivo della meridiana di s. Petronio, il fatto che il moto del Sole attorno alla Terra, quanto al movimento in declinazione, potesse essere rappresentato come un moto circolare attorno all'angolo dell'equinozio di ampiezza (oggi) di 23° 26', che è il valore dell'obliquita.
Il moto circolare proiettato sul meridiano della sfera celeste produce un moto armonico, di massima velocità agli equinozi e inversione del moto ai tropici (in greco la parola "tropico" vuol dire proprio "inversione").

L'equazione del moto armonico F=kx, nel punto di inversione con x massimo, descrive un moto accelerato che parte da fermo, e l'accelerazione tende a diminuire man mano che si muove, ma nelle immediate vicinanze della massima elongazione x l'accelerazione può considerarsi costante.

Quando l'accelerazione è costante la velocità cresce al trascorrere del tempo linearmente e lo spazio cresce col quadrato del tempo.
Questo è proprio quanto osservato nella tabella seguente, dove lo "spazio" è misurato in secondi d'arco dalla posizione zero corrispondente al transito del 21 dicembre.

| Data | z* del centro del Sole | Δz* del centro del Sole in arcsec ["] | | | Δz*da effemeridi IMCCE del centro del Sole ["] | | |
|---|---|---|---|---|---|---|---|
| 21 Dicembre | 65,33766 | +0.0 | | | + 0.0 | | |
| 22 Dicembre | 65,33669 | +3.5 | | | + 7.9 | | |
| 23 Dicembre | 65,32626 | +41.0 | :4 | 10.25 | +43.9 | :4 | 10.98 |
| 24 Dicembre | 65,30397 | +121.3 | :9 | 13.48 | +108.0 | :9 | 12.00 |
| 25 Dicembre | 65,28109 | +203.7 | :16 | 12.73 | +200.5 | :16 | 12.53 |
| 29 Dicembre | 65,101829 | +849.0 | :64 | 13.27 | +852.1 | :64 | 13.31 |

il valore dell'accelerazione (con il tempo misurato in giorni e lo spazio in secondi d'arco) non viene costante poiché l'origine del tempo non corrisponde esattamente con il solstizio, che è avvenuto alle 17:27 del 21 dicembre.
Il rapporto tra Δz e tempo al quadrato tende al valore vero dell'accelerazione con il crescere del tempo, anche se questo non è contato dalla vera origine, dunque assumo l'accelerazione a=12.73 (arcsec/di^2) e risolvo l'equazione 41=(2-s)^2*12.73, che è una misura più attendibile di 121.3" dalle considerazioni fatte prima.
L'equazione porta a s=0.205 giorni, ovvero 4.93 ore dopo il transito del 21 dicembre, cioè le 17:04, molto vicino al valore vero.

Una procedura di fit grafico permette di sfruttare meglio tutti i dati misurati per trovare le coordinate dell'origine della parabola di equazione Δz*=a*(t-s)^2-δ con δ l'angolo di cui il Sole doveva ancora scendere al transito meridiano del 21 dicembre.

Qui si segue una procedura matematica basata sulla soluzione di un sistema di equazione per sostituzioni successive.
Sono 3 incongnite a, s e δ ed abbiamo 6 equazioni escludendo quella del 18 dicembre in cui il foro aveva le dimensioni e la collocazione precedente a quella del 20 dicembre 2017 (Sigismondi, 2015 e video https://youtu.be/IKmOBj0kxgY ).

Scrivendo esplicitamente tutte le equazioni si vede come risolvere il sistema matematicamente, in particolare il termine as^2-δ è sempre presente in tutte le equazioni e vale zero, per cui lo scrivo solo nell'equazione del 21 dicembre. Da tutte le altre ricavo a ed s ed i relativi errori, e da quella del 21 il valore di δ.

[18 dicembre     Δz*=a*(-3-s)^2-δ=9a+6as=138.3"]
21 dicembre     Δz*=a*(0-s)^2-δ*=as^2-δ=0.0"
22 dicembre     Δz*=a*(1-s)^2-δ*=a-2s=3.5"
23 dicembre     Δz*=a*(2-s)^2-δ*=4a-4as=41.0"
24 dicembre     Δz*=a*(3-s)^2-δ*=9a-6as=121.3"
25 dicembre     Δz*=a*(4-s)^2-δ*=16a-8as=203.7"
29 dicembre     Δz*=a*(8-s)^2-δ*=64a-16as=849.0"

Sottraendo all'equazione del 29 il doppio di quella del 25 ottengo a=13.8
Sostituendo il valore di a nell'equazione del 23 ottengo s=0.257 giorni che valgono 6 ore e 10 minuti.
Sommandolo a 12h 08 si ottengono le 18h 18, in ottimo accordo con le effemeridi IMCCE che danno il solstizio per le 17h27.

Sostituendo nell'equazione del 21 a ed s ottengo δ=0.91".
Le equazioni del 18, 22 e 24 dicembre non le uso perché contengono più errori osservativi delle altre.
Trovato δ si ha anche la possibilità di conoscere l'obliquità ε, ovvero l'inclinazione dell'asse terrestre sul piano dell'orbita, che è la declinazione del Sole al momento esatto del solstizio.
Alla distanza zenitale del 21 dicembre aggiungiamo δ=0.91" e otteniamo quella che avrebbe avuto il Sole al solstizio.
z=65,33766°+0,91/3600=65,337914° che corrisponde ad una altezza sull'orizzonte h=90°-z=24,662086°

La declinazione è legata alla latitudine e all'altezza meridiana del Sole dall'equazione seguente

90°-latitudine dell'osservatorio+declinazione del Sole=altezza meridiana del Sole sull'orizzonte corretta per l'atmosfera.

nel nostro caso il 21 dicembre vale

declinazione del Sole=24,662086°-90°+41,90311°=-23,434803°: la declinazione del Sole al solstizio d'inverno è negativa.

La declinazione topocentrica al solstizio vale δ= -23° 26' 05.3".
Se il Sole fosse all'orizzonte la declinazione topocentrica coinciderebbe con quella geocentrica, poiché è a 24.66° di altezza occorre correggere per Raggio Terrestre*sen(24.66°)/Distanza Terra Sole istantanea= 6378*sen(24.66)°/147170762=0.065"
L'obliquità vera epsilon= -23° 26' 05.6" dal sito http://www.neoprogrammics.com/obliquity_of_the_ecliptic/
per il 21 dicembre h 17:27 basato sui calcoli di Jacques Laskar (Astronomy and Astrophysics 157,59 (1986) e sulle serie di nutazione IAU1980 e IAU2000B.
Attualmente la nutazione in epsilon vale -7.4"secondo l'andamento delle serie IAU1980 e 2000B ed il suo valore va sottratto all'obliquità media calcolata da Laskar per ottenere l'obliquità vera (inclusiva cioè della nutazione).
La nostra misura della declinazione del Sole al solstizio con i dati presi sulla meridiana di Santa Maria degli Angeli ci permette di misurare l'obliquità vera dell'eclittica entro 0.3" dal valore vero, ottenuto mediante l'uso combinato (la media) di molte osservazioni di astronomia di posizione fatte in osservatori specializzati per questo scopo e con strumenti moderni.
In Italia abbiamo il Centro di Geodesia Spaziale dell'Agenzia Spaziale Italiana di Matera dedicato a queste misure, mentre prima l'Osservatorio Astronomico di Carloforte sull'isola di S. Pietro in Sardegna era deputato all'International Latitude Service con misure di precisione sulla variazione della latitudine del luogo a 39°.08 fatte proprio con osservazioni stellari di questo tipo.

Si noti che con una precisione di 3"-4" è possibile misurare anche l'effetto della nutazione dell'asse terrestre, che vale +/-9".
Le misure fatte a santa Maria degli Angeli tengono conto di tutti questi effetti combinati insieme, che Bianchini non conosceva, poiché la nutazione è stata scoperta da Bradley solo nel 1737, così come l'aberrazione della luce scoperta dallo stesso nel 1727, che è la responsabile della misura di 41° 45' 30" della latitudine del foro stenopeico fatta nel 1702 da Bianchini osservando la Polare la cui luce era spostata dall'aberrazione (Sigismondi, 2009).

Si noti ancora che lo spostamento del foro di 2.6 mm verso Nord produrrebbe uno spostamento sulla linea meridiana verso Nord dell'immagine della stessa quantità. Con 1104 mm corrispondenti a 1083 mm dempta penumbra e 1950.89" abbiamo per ogni millimetro 1.8" di spostamento in angolo. Quindi con lo spostamento verso Nord di 2.6 mm avremo 4.7" in meno nel risultato sull'obliquità, compatibile comunque con la turbolenza atmosferica.

**Conclusioni: l'errore di misura alla meridiana di Bianchini da misure differenziali**

Ancora una volta le osservazioni del Sole attorno al solstizio d'inverno consentono di avvicinarsi ai fenomeni che furono scoperti nella prima metà del settecento, in astrometria stellare: si tratta dell'aberrazione della luce e della nutazione dell'asse terrestre. La meridiana di santa Maria degli Angeli consente di rilevare gli effetti di questi fenomeni in astrometria solare, e senza uso di ottiche rifrattive, ma solo del foro stenopeico, che è semplicemente vuoto.

La correzione di Cassini 60"*tan(z) è più che sufficiente per ottenere la precisione desiderata.
Questa misura è simile a quella condotta da Bianchini nel 1701 e comunicata a papa Clemente XI per lettera (Andrei, Sigismondi e Regoli, 2015), solo che oggi la meridiana non conserva più le tacche originali eccetto ai numeri 217, 213 e 212 e la maggiore incertezza nei risultati di oggi è dovuta proprio a questa ragione.

La rifrazione differenziale di 6" per il Sole meridiano al solstizio invernale di Roma non si può misurare dalle posizioni dei lembi, in quanto queste sono molto sensibili al contrasto luminoso che si ha in chiesa al momento della misura.
Il diametro meridiano dai dati decurtati della penombra risulta 1920" con 20" di deviazione standard, inferiore di 24" al valore reale, ma quello orizzontale, poiché l'immagine è in movimento, soffre maggiormente i problemi di contrasto luminoso e risulta più piccolo (1900") una volta decurtato della penombra (89").

La turbolenza atmosferica diurna di 3-4 secondi d'arco costituisce un limite per queste osservazioni, come lo costituirebbe per telescopi ottici più raffinati nella stessa posizione. Il contrasto luminoso dell'immagine non impedisce misure così accurate, purché i dati vengano corretti per la rifrazione atmosferica.
Le posizioni dei lembi meridionale e settentrionale del Sole consentono di ricavare la posizione del centro e conseguentemente la declinazione del solstizio invernale con una precisione corrispondente al seeing.
Alla meridiana di santa Maria degli Angeli l'obliquità vera dell'eclittica del solstizio invernale 2017 è stata misurata con una differenza inferiore al secondo d'arco sui valori tabulati secondo gli algoritmi IAU2000B e IAU1980.

**Bibliografia:**

Effemeridi IMCCE

```
# Body, Date, Rise, Azimut, Transit, Elevation, Set, Azimut, [Dawn-astronomical, Dawn-nautical, Dawn-civil, Twilight-civil, Twilight-nautical, Twilight-astronomical]
Sun, 2017-12-18, 07:34:05.9817, -58.4074, 12:06:41.6096, 24.7014, 16:39:14.6214, 58.3935, [05:51:50, 06:25:48, 07:00:59, 17:12:21, 17:47:32, 18:21:30]
Sun, 2017-12-19, 07:34:41.9261, -58.3743, 12:07:11.3144, 24.6800, 16:39:38.8959, 58.3647, [05:52:25, 06:26:23, 07:01:35, 17:12:46, 17:47:58, 18:21:56]
Sun, 2017-12-20, 07:35:15.8419, -58.3525, 12:07:41.1265, 24.6665, 16:40:05.4151, 58.3472, [05:52:58, 06:26:56, 07:02:08, 17:13:13, 17:48:25, 18:22:23]
Sun, 2017-12-21, 07:35:47.6861, -58.3421, 12:08:11.0082, 24.6607, 16:40:34.1464, 58.3411, [05:53:29, 06:27:28, 07:02:40, 17:13:42, 17:48:54, 18:22:53]
Sun, 2017-12-22, 07:36:17.4175, -58.3430, 12:08:40.9218, 24.6629, 16:41:05.0548, 58.3463, [05:53:59, 06:27:58, 07:03:10, 17:14:13, 17:49:25, 18:23:23]
Sun, 2017-12-23, 07:36:44.9979, -58.3553, 12:09:10.8299, 24.6729, 16:41:38.1035, 58.3630, [05:54:27, 06:28:26, 07:03:38, 17:14:46, 17:49:57, 18:23:56]
Sun, 2017-12-24, 07:37:10.3915, -58.3790, 12:09:40.6956, 24.6907, 16:42:13.2535, 58.3910, [05:54:54, 06:28:52, 07:04:03, 17:15:20, 17:50:32, 18:24:30]
Sun, 2017-12-25, 07:37:33.5654, -58.4141, 12:10:10.4823, 24.7164, 16:42:50.4643, 58.4303, [05:55:19, 06:29:17, 07:04:27, 17:15:57, 17:51:08, 18:25:05]
Sun, 2017-12-26, 07:37:54.4898, -58.4605, 12:10:40.1543, 24.7499, 16:43:29.6935, 58.4810, [05:55:42, 06:29:39, 07:04:49, 17:16:35, 17:51:45, 18:25:42]
Sun, 2017-12-27, 07:38:13.1383, -58.5182, 12:11:09.6768, 24.7913, 16:44:10.8976, 58.5430, [05:56:03, 06:29:60, 07:05:09, 17:17:15, 17:52:24, 18:26:21]
Sun, 2017-12-28, 07:38:29.4880, -58.5872, 12:11:39.0162, 24.8405, 16:44:54.0314, 58.6163, [05:56:23, 06:30:19, 07:05:27, 17:17:57, 17:53:05, 18:27:01]
Sun, 2017-12-29, 07:38:43.5200, -58.6674, 12:12:08.1404, 24.8974, 16:45:39.0491, 58.7008, [05:56:41, 06:30:35, 07:05:42, 17:18:40, 17:53:47, 18:27:42]
Sun, 2017-12-30, 07:38:55.2193, -58.7587, 12:12:37.0189, 24.9621, 16:46:25.9038, 58.7964, [05:56:56, 06:30:50, 07:05:56, 17:19:26, 17:54:31, 18:28:25]
```

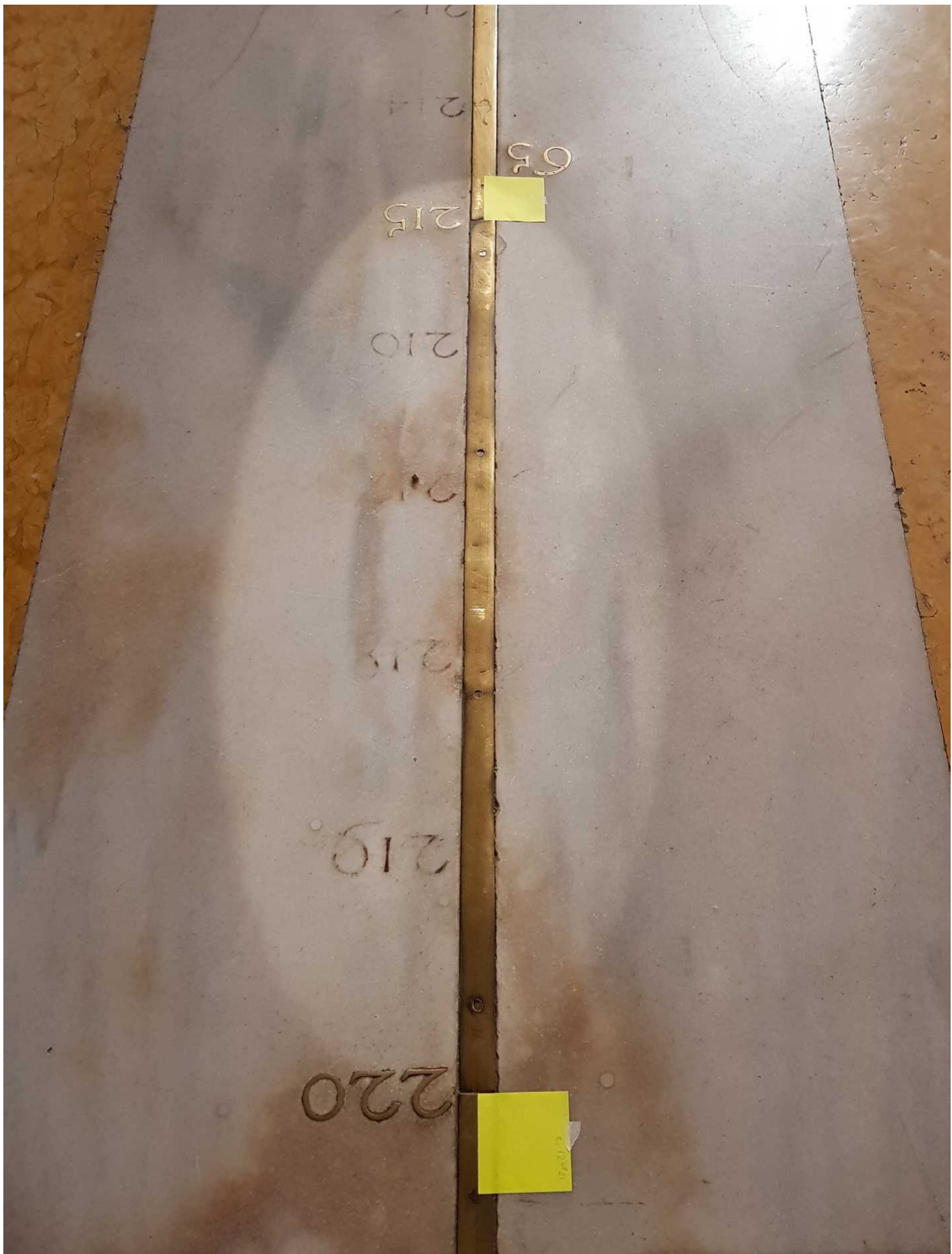

Immagine Solstiziale del 23 dicembre durante le misure (foto Enrico Giuliani)